\documentclass[prx,twocolumn, superscriptaddress]{revtex4-1}
\usepackage[dvipdfm, colorlinks, citecolor=blue]{hyperref}
\usepackage{epsfig}
\usepackage{graphicx}   
\usepackage{dcolumn}    
\usepackage{bm}         
\usepackage{color}
\usepackage{amssymb,amsmath}
\usepackage{txfonts}
\newcommand{\rr}{\mathbf{r}}
\newcommand{\lC}{\langle}
\newcommand{\rC}{\rangle}
\newcommand{\kk}{\mathbf{k}}

\begin{document}

\title{Parametrization of LSDA+$U$ for noncollinear magnetic configurations: \\
Multipolar magnetism in UO$_2$}
\author{S. L. Dudarev}
\email{sergei.dudarev@ukaea.uk}
\affiliation{UK Atomic Energy Authority, Culham Centre for Fusion Energy, Oxfordshire OX14 3DB, United Kingdom}
\affiliation{Department of Physics and Thomas
Young Centre, Imperial College London, South Kensington Campus,
London SW7 2AZ, United Kingdom}
\author{P. Liu}
\affiliation{University of Vienna, Faculty of Physics and Center for
Computational Materials Science, Sensengasse 8, A-1090 Vienna, Austria}
\author{D. A. Andersson}
\affiliation{Materials Science and Technology Division, Los Alamos National Laboratory, Los Alamos, New Mexico 87545, USA}
\author{C. R. Stanek}
\affiliation{Materials Science and Technology Division, Los Alamos National Laboratory, Los Alamos, New Mexico 87545, USA}
\author{T. Ozaki}
\affiliation{Institute for Solid State Physics, The University of Tokyo, Kashiwa 277-8581, Japan}
\author{C. Franchini}
\email{cesare.franchini@univie.ac.at}
\affiliation{University of Vienna, Faculty of Physics and Center for
Computational Materials Science, Sensengasse 8, A-1090 Vienna, Austria}
\affiliation{Dipartimento di Fisica e Astronomia, Universit\`{a} di Bologna, 40127 Bologna, Italy}

\begin{abstract}
To explore the formation of noncollinear magnetic configurations in materials with strongly correlated electrons, we derive a noncollinear LSDA+$U$ model involving only one parameter $U$, as opposed to the difference between the Hubbard and Stoner parameters $U-J$. Computing $U$ in the constrained random phase approximation,  we investigate noncollinear magnetism of uranium dioxide UO$_2$ and find that the spin-orbit coupling (SOC) stabilizes the 3$\kk$ ordered magnetic ground state. The estimated SOC strength in UO$_2$ is as large as 0.73 eV per uranium atom, making spin and orbital degrees of freedom virtually inseparable. Using a multipolar pseudospin Hamiltonian, we show how octupolar and dipole-dipole exchange coupling help establish the 3$\kk$ magnetic ground state with canted ordering of uranium $f$-orbitals. The cooperative Jahn-Teller effect does not appear to play a significant part in stabilizing the noncollinear 3$\kk$ state, which has the lowest energy even in an undistorted lattice. The choice of parameter $U$ in the LSDA+$U$ model has a notable quantitative effect on the predicted properties of UO$_2$, in particular on the magnetic exchange interaction and, perhaps trivially, on the band gap: The value of $U=3.46$ eV computed fully \emph{ab initio} delivers the band gap of 2.11~eV in good agreement with experiment, and a balanced account of other pertinent energy scales.
\end{abstract}

\maketitle

\section{Introduction}
\label{sec:01}

Predicting magnetic properties of complex materials by {\it ab initio} simulations requires using models that do not constrain the orientation of magnetic moments to a specific direction. This is achieved using noncollinear magnetic density functional theory approximations~\cite{Kubler1988,Sandratskii1998}, where the direction of local moments varies from point to point in real space. The fact that magnetic noncollinearity does occur in real materials is confirmed by experimental observations and {\it ab initio} calculations~\cite{Yamada1970, Heinze2000, Ohldag2001, HEIDE20092678, Hobbs2000, Hauptmann2017}.
Non-collinear magnetic ordering is particularly evident in compounds characterized by strong spin-orbit coupling (SOC) effects. Examples include $f$-electron systems~\cite{Faber1976,BURLET1986121,GIANNOZZI198775} and 5$d$ transition metal oxides~\cite{Jackeli2009,Liu2015}.
Non-collinear magnetic fluctuations contribute to electric and thermal resistivity of alloys~\cite{Starikov2018} and influence electronic and magnetic phase transitions~\cite{Kim2016,Kim2017a}. Magnetic fluctuations are also responsible for the anomalous thermal conductivity of uranium dioxide~\cite{Gofryk2014}, a most commonly used nuclear fuel.

In metallic alloys, magnetic fluctuations represent one of the modes of electronic excitations, contributing to electric and thermal resistivity. In a semiconducting oxide like UO$_2$, where thermal conductivity is dominated by phonons, the strong SOC couples atomic displacements with magnetic degrees of freedom and provides an additional channel of dissipation \cite{Gofryk2014}. The fact that lattice and magnetic degrees of freedom in uranium dioxide are not independent is confirmed by observations of piezo-magnetism~\cite{Jaime2017}. In Ref. \cite{Jaime2017} experimental observations were interpreted phenomenologically, assuming a direct coupling between atomic displacements and magnetic moments. At the electronic level, the effect stems from  the relativistic SOC involving magnetic moments of uranium ions and their orbital degrees of freedom. The directional character of bonds associated with $f$-orbitals generates inter-atomic forces that depend on the orientation of magnetic moments.

A magnetic metal and an actinide oxide like UO$_2$ differ in that magnetic orientation-dependent forces in a metal stem from the position-dependent Heisenberg's exchange~\cite{MaDu2016}, whereas in an actinide oxide the forces result from a combination of Anderson's superexchange~\cite{Anderson1950,Anderson1959}, strong correlations between electrons in $f$ shells~\cite{Kotani1992}, and relativistic spin-orbit interactions~\cite{Mironov2003,Savrasov2014a,Savrasov2014}.

In this paper, we explore the electronic structure and noncollinear magnetism of UO$_2$ using a suitably adapted {\it ab initio} LSDA+$U$ model. An {\it ab initio} treatment of directional magnetic degrees of freedom in a material with strong electron correlations requires a noncollinear model, which at the same time must be suitable for the evaluation of the total energy of the electronic system. An LSDA+$U$ model, often used for total energy calculations~\cite{Dudarev1998}, was derived from a Hamiltonian similar to the Hamiltonians used earlier by Anisimov {\it et al.}~\cite{Anisimov1991PRB}, and by Kotani and Yamazaki~\cite{Kotani1992}. These Hamiltonians are identical to the Hamiltonian of the collinear Stoner model~\cite{NguyenManh2009}, and hence are not suitable for the treatment of noncollinear magnetism. A noncollinear magnetic LSDA+$U$ model was proposed by Liechtenstein {\it et al.}~\cite{Liechtenstein1995} and Solovyev {\it et al.} \cite{Solovyev1998}. In Ref. \cite{Liechtenstein1995} the choice of the double counting term, which has the same form as in the collinear case, {\it cf.} Eq.~(4) of Ref.~\cite{Liechtenstein1995} and Eq.~(3) of Ref.~\cite{NguyenManh2009}, introduces an element of uncertainty in the total energy part of the analysis. The mean-field treatment developed in Ref. \cite{Solovyev1998} involved the entire model Hamiltonian, making it difficult to separate the LSDA+$U$ treatment of correlations from the exchange-correlation terms already included in the density functional theory (DFT) approximation. Various forms of the  LSDA+$U$ model were investigated in Refs. \cite{Anisimov1997,Cococcioni2012,Mazurenko2007,Ylvisaker2009,Bousquet2010,Mellan2015}.

This study is based on a recent analysis by Coury {\it et al.}~\cite{Coury2016,*Coury2016_Erratum}, who found a way of transforming, through an exact calculation, the general second-quantized Hamiltonian for interacting electrons into a form similar to that used in earlier LSDA+$U$ studies \cite{Anisimov1991PRB,Dudarev1998}. Coury {\it et al.} ~\cite{Coury2016,*Coury2016_Erratum} showed that the model Hamiltonians used earlier were missing a term  contributing to the LSDA+$U$ correction. The inclusion of this term, as we show below, simplifies the LSDA+$U$ equations and enables performing an effectively parameter-free DFT study, where parameter $U$ is deduced from a constrained random phase approximation (RPA) calculation.

Uranium dioxide, extensively studied in Refs. \cite{Allen1968a,Allen1968b,BURLET1986121,Caciuffo1999,Laskowski2004,Zhou2011,Korotin2016,Pegg2017,Bruneval2018,Pegg2019a,Pegg2019b,PhysRevB.82.241103,PhysRevB.88.195146},
continues attracting considerable attention as the scope of {\it ab initio} methods expands to enable the treatment of defects~\cite{Crocombette2001} and their diffusion~\cite{Dorado2012,Vathonne2017,Liu2018a}, given that defects and their diffusion contribute to the
overall performance of UO$_2$ nuclear fuel. The magnetic ground state of bulk crystalline UO$_2$ has been investigated experimentally and theoretically~\cite{Allen1968a,Allen1968b,BURLET1986121,Caciuffo1999,Laskowski2004,Zhou2011,Caciuffo2011,PhysRevB.88.195146}, and there is extensive information about the surface structure of UO$_2$ derived from elevated temperature scanning tunneling microscope observations \cite{Castell1998a,Castell1998b,Castell1999}. There is still no definitive verdict about what stabilizes its 3$\kk$ magnetic ground state {\it and} determines the spectrum of low-energy magnetic excitations required, for example, for spin-lattice dynamics simulations~\cite{Ma2008,MaDu2016,Tranchida2018}. Partly, the difficulty stems from the fact that an {\it ab initio} treatment of UO$_2$ requires exploring an energy landscape with multiple local minima, which is difficult to treat using conventional energy minimization algorithms~\cite{Dorado2013}. It is also necessary to take into account relativistic effects, giving rise to large SOC~\cite{Korotin2016} and multipolar spin interactions. In UO$_2$, magnetism is associated with uranium $f$ orbitals, with the orbital magnetic moments of uranium ions being twice the spin moments~\cite{Dudarev2000}. Magnetic order is also believed to be linked to the Jahn-Teller (JT) lattice distortions~\cite{Sasaki1970}.

Presently available LSDA+$U$ models require using values of $U$ and $J$ as input parameters. These quantities are often treated as being tunable, and are matched to the calculated band gap, the equilibrium volume, the magnetic moment or the formation energy~\cite{Franchini2007}. Several approaches have been proposed to compute $U$ and $J$ and avoid using phenomenological considerations~\cite{PhysRevB.74.125106,Cococcioni2005,Solovyev2006,Anisimov1991}. The constrained random phase approximation (cRPA) accounts for the screening of Coulomb interaction between correlated electrons and provides estimates for $U$ and $J$ in a constrained correlated subspace. cRPA has been applied to a variety of materials, and enabled obtaining fairly accurate values of $U$ and $J$~\cite{Vaugier2012,Kim2017,Kim2018,Liu2018}. In the analysis below, we use cRPA to compute $U$ and $J$, in this way enabling a parameter-free LSDA+$U$ simulation. The approach based on Coury's Hamiltonian \cite{Coury2016,*Coury2016_Erratum} requires computing only one parameter $U$.

Below, we derive a noncollinear LSDA+$U$ model, providing equations for the effective one-electron potential and the double counting correction to the total energy, and find that the model requires only one parameter $U$ as opposed to the difference $U-J$ entering the existing LSDA+$U$ equations. cRPA is used for computing $U$. The stability of the noncollinear 3$\kk$ magnetic ground state in UO$_2$ is investigated using an adiabatic occupation matrix approach. Analysis shows that the 3$\kk$ structure represents the lowest energy configuration even in an undistorted cubic lattice. Finally, we discuss effective magnetic Hamiltonians for finite temperature atomic and magnetic dynamic simulations.

\section{a Non-collinear LSDA+$U$ model}
\label{sec:02}
An LSDA+$U$ model aims to provide an improved description of the electronic structure of materials characterized by strong electron correlations in spatially localized {\it d} and {\it f} shells. This is achieved by adding a correction term to the effective single particle electron potential~\cite{Anisimov1991PRB,Liechtenstein1995,Dudarev1998},
\begin{equation}
V^{\sigma}_{jl}={\delta E_{{\rm LSDA}+U}\over \delta \rho ^{\sigma}_{lj}}={\delta E_{\rm LSDA}\over \delta \rho ^{\sigma}_{lj}}+
({ U}-{ J})\left[{1\over 2}\delta_{jl}-\rho ^{\sigma}_{jl}\right],\label{V}
\end{equation}
and a double counting correction to the total electronic energy~\cite{Dudarev1998},
\begin{equation}
E^{dc}_{{\rm LSDA}+U}={(U-J)\over 2}\sum _{\sigma, j,l}\rho ^{\sigma}_{jl}\rho ^{\sigma}_{lj}.\label{DC}
\end{equation}
The latter is necessary since a  sum of single particle energies of interacting electrons does not represent their total energy; see Eqs.~(15) and (16) of Ref.~\cite{Dudarev2007}. In Eqs.~(\ref{V}) and (\ref{DC}), indexes $j,l$ refer to the orbitals associated with a lattice site, and $\sigma$ is the spin index.

Equations~(\ref{V}) and (\ref{DC}) were derived from a model tight-binding Hamiltonian~\cite{Caroli1969,Kotani1992}, where the on-site electron interaction terms have the form,
\begin{equation}
\hat {\cal H}={U\over 2}\sum _{l,l',\sigma}\hat n_{l,\sigma} \hat n_{l',-\sigma}+{(U-J)\over 2}\sum _{l,l',l\ne l',\sigma}\hat n_{l,\sigma} \hat n_{l',\sigma}.\label{H_collinear}
\end{equation}
It can be shown~\cite{NguyenManh2009,Coury2016} that this Hamiltonian is identical to the Hamiltonian of the collinear Stoner model,
\begin{equation}
\hat {\cal H}={U\over 2}\left(\hat N^2-\hat N\right)-{J\over 4}\left(\hat N^2-2\hat N\right)-{J\over 4}\hat M^2,\label{Stoner_collinear}
\end{equation}
where $\hat N_{\sigma}=\sum _l\hat n_{l,\sigma}$, $\hat N=\hat N_{\uparrow}+\hat N_{\downarrow}$, and $\hat M=\hat N_{\uparrow}-\hat N_{\downarrow}$. Equations (\ref{V}) and (\ref{DC}) can be derived by evaluating the expectation values of either Eq.~(\ref{H_collinear}) or Eq.~(\ref{Stoner_collinear}); see Ref.~\cite{Dudarev1998}.

Despite the relative success of the LSDA+$U$ model~\cite{Cococcioni2012}, there are two points that require attention. It is unclear to what extent the choice of the model Hamiltonian (\ref{H_collinear}) affects the form of Eqs.~(\ref{V}) and (\ref{DC}), and also how to generalize these equations to noncollinear magnetic configurations~\cite{Zhou2011,Gofryk2014,MaDu2016}.

First, we note that there is a significant term missing in Hamiltonians (\ref{H_collinear}) and (\ref{Stoner_collinear}). This missing term has been identified in Ref.~\cite{Coury2016,*Coury2016_Erratum}. This term contributes to the LSDA+$U$ correction, changing its form. Second, we note that the correction itself is not invariant in the extended space of spin and orbital indexes, a point that can be readily rectified using a suitable definition of the convolution of the full spin- and orbital-dependent electron density matrix. The required invariant form was proposed in \cite{Rohrbach2003} and already implemented in VASP,  although the coefficients used in the numerical implementation were chosen as in Eqs.~(\ref{V}) and (\ref{DC}). This now requires modification, as we show below.

An on-site Hamiltonian, describing interaction between electrons occupying orbitals $i,j,k,l$, is given by a sum of combinations of four creation and annihilation operators multiplied by a four-index matrix $V_{ij,lk}$:
\begin{equation}
\hat {\cal H}=\frac{1}{2}\sum_{i,j,k,l}\sum_{\sigma\xi}V_{ij, lk} \hat{c}^{\dagger}_{i,\sigma}\hat{c}^{\dagger}_{j,\xi}\hat{c}_{k,\xi}\hat{c}_{l,\sigma}.\label{H_full}
\end{equation}
Matrix $V$ has (2$l$+1)$^4$ elements, which in the case of {\it p} ($l$=1) electrons amounts to $3^4$=81 elements, and $5^4$=625 elements in the case of {\it d} electrons. Symmetry constraints show that all the elements of $V_{ij,lk}$ can be parameterized using only two independent constants in the {\it p}-electron case, three constants in the {\it d}-electron case, and four in the {\it f}-electron case. In the {\it p}-electron cubic harmonic orbital case, using an analogy with the theory of isotropic elasticity~\cite{Kosevich2005}, where the four-index matrix of elastic constants $C_{ijkl}$ has the same symmetry as $V_{ij,lk}$, Hamiltonian (\ref{H_full}) can be written {\it exactly} as~\cite{Coury2016,*Coury2016_Erratum}
\begin{equation}
\hat {\cal H}={1\over 2}\left(U-{J\over 2}\right):\hat N^2:-{J\over 4}:\hat{\bm M}^2:+{J\over 2}\sum _{i, j} :(\hat n_{ij})^2:.\label{H_Coury}
\end{equation}
Here, $\hat N=\sum_{m,\sigma}\hat c^{\dagger}_{m, \sigma}\hat c_{m, \sigma}$ is the operator of the total number of electrons on a site, $\hat n_{kl}=\sum_{\sigma}\hat c^{\dagger}_{k, \sigma}\hat c_{l, \sigma}$, and $$\hat {\bf M}=\sum _{m, \xi, \xi'} \hat c^{\dagger}_{m, \xi}{\bm \sigma}_{\xi\xi'}\hat c_{m, \xi'}$$ is the total magnetic moment vector operator associated with a site. In Eq.~(\ref{H_Coury}), :: denotes normal ordering of creation and annihilation operators, and ${\bm \sigma}_{\xi\xi'}$ are the Pauli matrices. The normally ordered terms in Eq.~(\ref{H_Coury}), expressed using conventional notations, have the form,
\begin{eqnarray}
:\hat N^2:&=&\hat N^2-\hat N, \nonumber \\
:\hat {\bf M}^2:&=&\hat {\bf M}^2-3\hat N,\nonumber \\
:\hat M_z^2:&=&\hat M_z^2-\hat N,\nonumber \\
:(\hat n_{kl})^2:&=&\sum_{\sigma,\xi}\hat c^{\dagger}_{k,\sigma}\hat c^{\dagger}_{k,\xi}\hat c_{l,\xi}\hat c_{l,\sigma}.\label{N_ordered}
\end{eqnarray}
The first two terms in Eq.~(\ref{H_Coury}) are the same as the right-hand side of  Eq.~(\ref{Stoner_collinear}), with the exception that now the magnetic moment operator is a vector quantity. In addition, Hamiltonian (\ref{H_Coury}) includes an extra third term, required by symmetry and absent in Eq.~(\ref{Stoner_collinear}). This term is related to the orbital moment of electrons on a lattice site~\cite{Coury2016,*Coury2016_Erratum}.  We note that Hamiltonian (\ref{H_Coury}) is exact in the sense that no procedure of ``directional'' averaging is involved in the transformation from Eq.~(\ref{H_full}) to Eq.~(\ref{H_Coury}). The central approximation associated with a Hamiltonian of the form (\ref{H_Coury}) is that it represents a subset of localized orbitals on an individual site taken in isolation, and does not include the fact that the self-consistent field of neighboring ions might influence its rotational invariance. Also, the mean-field approximation adopted in the treatment below, while being fairly well documented in the context of LSDA+$U$ models, requires critical assessment in applications.

We now follow the derivation given in Refs.~\cite{Anisimov1991PRB,Dudarev1998} and deduce the LSDA+$U$ model from Hamiltonian (\ref{H_Coury}). We identify the terms in Hamiltonian (\ref{H_Coury}) that contain two creation and two annihilation operators acting on the same electronic state $(m,\sigma)$.  In the mean-field approximation, these terms provide contribution to the total energy proportional to $n^2_{m,\sigma}$ whereas their exact expectation value is proportional to $n_{m,\sigma}$. The LSDA+$U$ model correction equals the difference between the exact and mean-field expectation values of these terms, resulting in
\begin{eqnarray}
&&E_{{\rm LSDA}+U}-E_{\rm LSDA}=\nonumber \\
&&\left[{1\over 2}\left(U-{J\over 2}\right)-{J\over 4}+{J\over 2} \right]\sum_{m,\sigma}\left(n_{m,\sigma}-n^2_{m,\sigma}\right).\label{new1}
\end{eqnarray}
In the above expression, each term in square brackets corresponds to a respective term in Hamiltonian (\ref{H_Coury}), and $n_{m\sigma}$ is the electron occupation number of an orbital state $m$ with spin index $\sigma$. The term $(J/2)\sum_{m,\sigma}\left(n_{m,\sigma}-n^2_{m,\sigma}\right)$, missing in the derivations given in Refs.~\cite{Anisimov1991PRB,Dudarev1998}, results from the last term in Eq.~(\ref{H_Coury}).

To illustrate the derivation of Eq.~(\ref{new1}), consider, for example, the term $-(J/4){\hat{\bf M}}^2$  in Eq.~(\ref{H_Coury}). In explicit form, this operator can be written as
$$
-{J\over 4}\hat{\bf M}^2=-{J\over 4}\sum _{\alpha , \xi, \xi '}\sum _{\beta , \zeta, \zeta '} (\hat c^{\dagger}_{\alpha \xi
}{\bm \sigma}_{\xi \xi '}\hat c _{\alpha \xi '})(\hat c^{\dagger}_{\beta \zeta
}{\bm \sigma}_{\zeta \zeta '}\hat c _{\beta \zeta '}).
$$
The part of the above operator expression where the indexes of all the creation and annihilation operators coincide, is
$$
-{J\over 4}\sum _{\alpha , \zeta} \hat c^{\dagger}_{\alpha \zeta
}\hat c _{\alpha \zeta})\hat c^{\dagger}_{\alpha \zeta
}\hat c _{\alpha \zeta }({\bm \sigma}_{\xi \xi }\cdot {\bm \sigma}_{\zeta \zeta }).
$$
Since $\sigma ^{x}_{\zeta \zeta }\sigma ^{x}_{\zeta \zeta }+\sigma ^{y}_{\zeta \zeta }\sigma ^{y}_{\zeta \zeta }+\sigma ^{z}_{\zeta \zeta }\sigma ^{z}_{\zeta \zeta }=1$, we see that the form of the operator expression that requires applying the LSDA+$U$ correction is the same as the operator form arising from the $\hat N^2$ term, and hence the contribution to the LSDA+$U$ functional from the $-(J/4){\hat{\bf M}}^2$ term in Hamiltonian (\ref{H_Coury}) equals
$$
-{J\over 4} \sum _{\alpha, \zeta}(n_{\alpha \zeta}- n ^2_{\alpha \zeta}).
$$
Applying this procedure to all the terms in (\ref{H_Coury}),  we see that the terms in the LSDA+$U$ correction Eq.~(\ref{new1})  that contain parameter $J$ cancel each other exactly, and only the term proportional to parameter $U$ remains. The sum of the first two terms in square brackets $(U-{J/2})/2-J/4=(U-J)/2$ equals the coefficient found in earlier derivations~\cite{Dudarev1998} based on Hamiltonian (\ref{Stoner_collinear}). The third term in square brackets in Eq.~(\ref{new1}) stems from the last term in Hamiltonian (\ref{H_Coury}), missing in (\ref{Stoner_collinear}). We note that the complete cancellation of the terms containing parameter $J$ in a derivation based on a full Hamiltonian (\ref{H_Coury}) should not come as a surprise. For example, the original form of the Hubbard Hamiltonian~\cite{Hubbard1963} contains no $J$ terms but still generates a variety of magnetic solutions, originating solely from strong on-site electron correlations. Furthermore, despite the appealing simplicity of Hamiltonian (\ref{H_collinear}), it was in fact never derived directly from the four-index matrix form (\ref{H_full}) and hence it is not unexpected that a direct derivation undertaken by Coury {\it et al.} \cite{Coury2016,*Coury2016_Erratum} showed that  Eqs.~(\ref{H_collinear}) and (\ref{H_full}) were not fully consistent.

A general form of Eq.~(\ref{new1}), invariant with respect to the choice of electronic orbitals and spin quantization axis, is
\begin{eqnarray}
&&E_{{\rm LSDA}+U}-E_{\rm LSDA}={U\over 2}\left[{\rm Tr}\rho-{\rm Tr}\rho^2\right]\nonumber \\
&=&{U\over 2}\left[\sum _{m,\sigma}\rho_{mm}^{\sigma\sigma}-\sum _{m,\sigma; m',\sigma'}\rho_{mm'}^{\sigma\sigma'}\rho_{m'm}^{\sigma'\sigma}\right],\label{new2}
\end{eqnarray}
where $\rho$ is the full orbital and spin-dependent one-electron density matrix.

In the collinear approximation, where the density matrix is diagonal with respect to the subset of its spin indexes $\rho_{mm'}^{\sigma\sigma'}=\rho_{mm'}^{\sigma}\delta_{\sigma\sigma'}$, Eq.~(\ref{new2}) is similar to Eq.~(5) of Ref.~\cite{Dudarev1998}, however, the prefactor in the formula is still different.

An invariant orbital- and spin-dependent noncollinear form of LSDA+$U$ (\ref{V}) and (\ref{DC}) is now:
\begin{equation}
V^{\sigma\sigma'}_{jl}={\delta E_{{\rm LSDA}+U}\over \delta \rho ^{\sigma\sigma'}_{lj}}={\delta E_{\rm LSDA}\over \delta \rho ^{\sigma\sigma'}_{lj}}+
U\left[{1\over 2}\delta_{jl}\delta_{\sigma\sigma'}-\rho ^{\sigma\sigma'}_{jl}\right],\label{V_new}
\end{equation}
and
\begin{equation}
E_{{\rm LSDA}+U}^{dc}={U\over 2}\sum _{\sigma,\sigma', j,l}\rho ^{\sigma\sigma'}_{jl}\rho ^{\sigma'\sigma}_{lj}.\label{DC_new}
\end{equation}
The need to add the double counting term (\ref{DC_new}) to the total energy, evaluated using the conventional Kohn-Sham procedure, stems from the fact that the single-particle electron potential (\ref{V_new}) depends on the occupancy of electron orbitals through a term proportional to $\rho ^{\sigma\sigma'}_{jl}$. It is this occupancy-dependent term in Eq.~(\ref{V_new}) that makes a sum of one-particle energies different from the total energy given by Eq.~(\ref{new2}). The above equations show that, in addition to correcting the prefactor in the formula, an invariant LSDA+$U$ model requires convoluting the density matrix over the full set of its orbital and spin indexes, a point that was not included in earlier derivations~\cite{Anisimov1991PRB,Liechtenstein1995,Dudarev1998}. The LSDA+$U$ correction of the form (\ref{new2}) and (\ref{DC_new}) has already been implemented in VASP \cite{Rohrbach2003}, but with the coefficients given by (\ref{V}) and Eq.~(\ref{DC}). The derivation above shows that in a practical calculation it is sufficient to set $J=0$ in the existing noncollinear implementation of the method \cite{Rohrbach2003} to arrive at the LSDA+$U$ correction consistent with the full model Hamiltonian (\ref{H_Coury}).

The terms containing parameter $J$ also cancel exactly if we perform the above derivation for the {\it d}-electron case~\cite{Coury2016,*Coury2016_Erratum}. The most direct way of showing this involves starting from Eq.~(22) of Ref.~\cite{Coury2016,*Coury2016_Erratum} and noting that in the {\it d}-electron case, all the terms containing parameter $J$ can be expressed in terms of a renormalized parameter $J-6\Delta J$, resulting in the same Eq.~(\ref{new2}) above, plus small terms proportional to $\Delta J$, which together amount to a small fraction of an electronvolt per atom and are normally neglected in applications \cite{Cococcioni2012}. This suggests that the single-parameter form of the LSDA+$U$ correction given by Eq.~(\ref{V_new}) and Eq.~(\ref{DC_new}) remains sufficiently accurate and applicable to {\it d}-electron orbitals, and other types of shells containing correlated electrons.  Parameter $U$, according to the analysis given in~\cite{Anisimov1997,Vaugier2012}, is an effective quantity, characterizing the strength of electron-electron interactions and modified by many-body self-screening.

Concluding this section and before proceeding to the {\it ab initio} analysis, we note that Eqs.~(\ref{new2}), (\ref{V_new}), and (\ref{DC_new}) amount to only a small correction to the established exchange-correlation functionals of density functional theory. The magnitude of the correction term (\ref{new2}) does not exceed $(U/8)$ times the number of partially filled orbitals, which in practical calculations amounts to approximately no more than one electron-volt per ion.

\section{\emph{Ab Initio} Methodology}

All the calculations below were carried out using the Vienna \emph{ab initio} simulation package (VASP)~\cite{PhysRevB.47.558,PhysRevB.54.11169}, where the noncollinear LSDA+$U$ scheme (\ref{new2}) and (\ref{DC_new}) is implemented~\cite{Bengone2000,Rohrbach2003} with the full inclusion of relativistic effects and self-consistent treatment of spin-orbit coupling~\cite{Hobbs2000}.
A robust energy cut off up to 700 eV with the convergence precision of 10$^{-6}$ eV was used in all the calculations, and the Brillouin zone was sampled using a 6$\times$6$\times$6 $k$-point mesh. Atomic positions were optimized with the lattice parameters fixed at its observed value ($a$=5.469 \AA)~\cite{MOMIN1991308}. Among the points that we explore in detail below are (A) the evaluation of interaction parameters $U$ and $J$ using the cRPA, (B) magnetically constrained DFT calculations, (C) spin adiabatic occupation matrix analysis of the magnetic energy landscape, (D) parametrization of the multipolar pseudospin Hamiltonian and exchange coupling, and (E) evaluation of the strength of SOC.

\subsection{Constrained random phase approximation}
\label{sec:crpa}
Interaction parameters $U$ and $J$ were computed from first principles using the constrained random phase approximation~\cite{PhysRevB.74.125106}. In the cRPA, the Coulomb repulsion and Hund's coupling parameters $U$ and $J$ are derived from the matrix elements of $U_{ijkl}$ written in terms of the Wannier basis functions, representing the correlated subspace (uranium $f$ states)
\begin{equation}
U_{ijkl}=\lim_{\omega \to 0}\iint d\rr d\rr' w^*_i(\rr) w^*_j(\rr') \mathcal{U}(\rr,\rr',\omega) w_k(\rr) w_l(\rr').
\label{eq:cRPA}
\end{equation}
$U$ and $J$ are the matrix elements $U_{ijij}$ and $U_{ijji}$, respectively. In Eq.~(\ref{eq:cRPA}), $\mathcal{U}$ is the partially screened interaction kernel, which is evaluated by solving the Dyson-like equation
\begin{equation}
\mathcal{U}^{-1}=\mathcal{V}^{-1}-\chi^r,
\label{eq:Uijkl_3}
\end{equation}
where $\mathcal{V}$ is the bare (unscreened) interaction kernel and $\chi^r = \chi-\chi^t$ is the polarizability, excluding contributions from the ``target'' correlated $f$ subspace, $\chi^t$.
Following the above procedure, we find $U^{\rm{cRPA}}$=3.46~eV and $J^{\rm{cRPA}}$=0.30~eV, corresponding to the effective interaction parameter $U_{\rm eff}^{\rm{cRPA}}$=3.16~eV. These values are smaller than those extracted from optical spectroscopic estimates \cite{Dudarev1997}:
$U$=4.50~eV and $J$=0.54~eV, $U_{\rm eff}$=3.96~eV.
In order to estimate the effect of the prefactor in Eq.~(\ref{new2}) on magnetic properties of UO$_2$, we have compared four different choices of interaction parameters, namely
\begin{enumerate}
\item $U$=3.16~eV ~~ (i.e. $U_{\rm eff}^{\rm{cRPA}}$)
\item $U$=3.46~eV ~~ (i.e. $U^{\rm{cRPA}}$)
\item $U$=3.96~eV ~~ (i.e. `Expt.' $U_{\rm eff}$)
\item $U$=4.50~eV ~~ (i.e. `Expt.' $U$)
\end{enumerate}

\begin{figure}
\includegraphics[width=0.45\textwidth,clip]{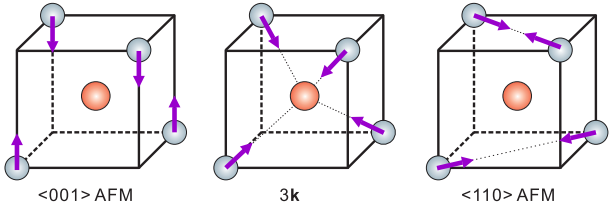}
\caption{Schematic plots of the $\langle{001}\rangle$-AFM, 3$\kk$ (longitudinal) and $\langle{110}\rangle$-AFM noncollinear spin configurations considered in this study.
}
\label{fig:models}
\end{figure}

\subsection{Magnetically constrained noncollinear DFT+$U$}

To model the noncollinear magnetic ground state of UO$_2$ we have minimized the total energy, treating it as a function of  directions of spin moments through magnetically constrained noncollinear DFT+$U$~\cite{Hobbs2000,Liu2015,Ma2015}. We have inspected spin rotations that transform the system from a characteristic noncollinear 3$\kk$ state into collinear antiferromagnetically (AFM) ordered $\langle 001\rangle $ and $\langle 110\rangle$ configurations~\cite{Zhou2011}, illustrated in Fig.~\ref{fig:models}. A noncollinear 3$\kk$ phase is described by three independent wave vectors and can be represented by a combination of three different phases, one longitudinal and two equivalent transverse. To facilitate the construction of the canted magnetic energy landscape, we used the longitudinal 3$\kk$ ordered magnetic structure shown in Fig.~\ref{fig:models} as a starting configuration. The two other ordered AFM configurations, $\langle{001}\rangle $ and $\langle{110}\rangle$, belong to the 1\textbf{k} (one wave-vector) and 2\textbf{k} (two wave-vectors) categories, respectively.

The $\langle{001}\rangle$--3$\kk$--$\langle{110}\rangle$ magnetic structure transformation pathway can be defined by a concerted variation of angle $\theta$ on the four inequivalent uranium sites in a UO$_2$ magnetic unit cell [see Figs.~\ref{fig:theta}(a) and~\ref{fig:theta}(c)]. Constrained energy minimization as a function of $\theta$ along the transformation pathway is achieved by considering the energy penalty arising from a constraint applied to the direction of the spin magnetic moment, defined by the function,
\begin{equation}
E= E_0(\{{\bf M}_i\})+\sum_{i}\gamma[\mathbf{M}_i-\mathbf{M}^0_i(\mathbf{M}^0_i\cdot\mathbf{M}_i)]^2.
\label{eq:cmag}
\end{equation}
Here $E_0$ is the unconstrained DFT total energy, whereas the second term is a penalty contribution defined as a noncollinear directional constraint on the direction of local moments $\mathbf{M}_i$ with respect to an arbitrary set of unit vectors ${\mathbf{M}^0_i}$ on sites $i$. $\mathbf{M}_i$ is the magnetic moment computed by integrating over a Wigner-Seitz cell centered on atom $i$ (the effective Wigner-Seitz radius is 1.588~\AA~for a U ion and 0.82~\AA~for an O ion). Parameter $\gamma$ defines the magnitude of the energy penalty term. By progressively increasing $\gamma$, functional (\ref{eq:cmag}) is driven to convergence towards the DFT total energy \cite{Ma2015}. We used the value of $\gamma$=10 eV$/\mu_B^2$ that guarantees that the expectation value of the energy penalty term at the energy minimum found through the application of the constrained minimization procedure (\ref{eq:cmag}) is lower than 10$^{-5}$ eV.

\subsection{Adiabatic spin occupation matrix approach}

\begin{figure}
\includegraphics[width=0.45\textwidth,clip]{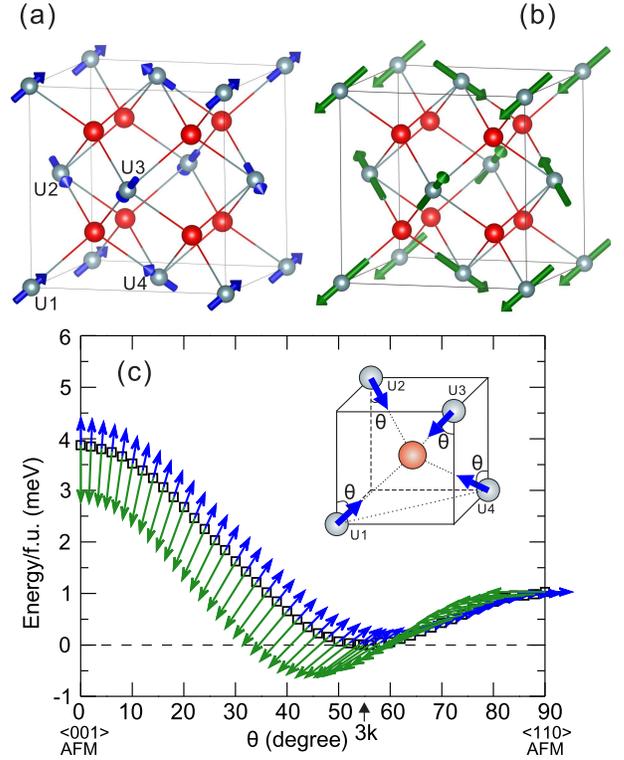}
\caption{
Schematic view of (a) spin and (b) orbital moments in a 3$\kk$ (longitudinal) magnetic unit cell of UO$_2$. Panel (c) shows the total energy as a function of the canting angle $\theta$ along the $\lC 001 \rC$-AFM -- 3$\kk$ -- $\lC 110 \rC$-AFM transformation pathway. Blue and green arrows show the spin and orbital moments, respectively. The inset explains the definition of the spin canting angle $\theta$  adopted in our analysis.}
\label{fig:theta}
\end{figure}

A known drawback of DFT+$U$ approaches is the difficulty associated with finding the lowest energy state of a strongly correlated magnetic material. In most cases a DFT+$U$ functional exhibits a multitude of local minima corresponding to a variety of spin and orbital occupancies in the correlated electronic   subspace~\cite{Dorado2009, Meredig2010, Zhou2011, Watson2014}.  The difficulty with finding a global minimum stems from the curvature of the energy surface as a function of orbital occupations \cite{Meredig2010}. In DFT calculations the energy surface is typically convex, but the global minimum might correspond to a physically unreasonable partial fractional orbital occupation predicting a metallic state of a material that in reality is an insulator. DFT+$U$ corrects this by adding a term that penalizes fractional occupations, but this correction changes the curvature of the energy surface from convex to concave, producing many local energy minima~\cite{Meredig2010}.
Dorado and coworkers addressed this point by performing a search involving a large number of self-consistent calculations, each starting from different initial occupation matrices, and selected the outcome corresponding to the lowest total energy~\cite{Dorado2009}. This procedure can be accelerated by adiabatically ``turning on" the value of parameter $U$ starting from the DFT limit $U=0$ and gradually converging to the true ground state with integral orbital occupations~\cite{Meredig2010}.
The above issue is particularly pertinent to noncollinear spin systems, where small rotations of spin moments could give rise to many local minima, all contained within a few meV energy interval.

Bearing in mind this aspect of energy minimization, we have combined the occupation matrix approach with a gradual adiabatic change of the spin moment direction, using the magnetically constrained noncollinear DFT+$U$ functional described in the previous subsection~\cite{Hobbs2000,Liu2015,Ma2015}.
Starting from the 3$\kk$-type noncollinear spin ordered state of UO$_2$~\cite{BURLET1986121,Zhou2011,Pegg2019a} shown in Fig.~\ref{fig:theta}(a), we gradually changed the canting angle $\theta$, moving adiabatically from the noncollinear 3$\kk$ state to the energetically comparable AFM collinearly ordered $\langle 001\rangle $ and $\langle 110\rangle $ configurations shown in Fig.~\ref{fig:theta}(c). At each canting step, we initialized the occupation matrix to the one obtained at the preceding step and performed a fully self-consistent calculation. In this way, by gradually perturbing the wave function of the 3$\kk$ state, we were able to construct a smooth total energy curve E($\theta$) as a function of the canting angle $\theta$, shown in Fig.~\ref{fig:theta}(c). The absence of cusps and sudden jumps guarantees that this energy curve represents the lowest energy path linking the spin configurations considered here, and also that the 3$\kk$ state is indeed the global minimum with respect to spin rotations. Probing other spin configurations (not shown here) confirmed the outcome of this analysis and resulted in spin configurations with the lowest total energies. We note that the orbital moment ${\bf m}_o$ remains anti-parallel to the spin moment ${\bf m}_s$ everywhere on the transformation pathway, and the magnitude of both moments remains almost independent of the canting angle: $m_s \approx 1.5~\mu_B$ and $m_o \approx 3.2~\mu_B$; see  Figs.~\ref{fig:theta}(b) and~\ref{fig:theta}(c).

\subsection{Effective pseudospin Hamiltonian and exchange interactions}

To characterize magnetic properties of a material and understand the origin of the specific spin ordered configuration that it adopts, it is necessary to quantify the dominant spin-spin interactions.  For systems conserving the total spin moment, magnetic coupling parameters can be analyzed in terms of an effective Heisenberg spin Hamiltonian involving conventional spin operators. In materials with  strong spin-orbit coupling, the spin moments alone are not conserved and it is more appropriate to use the pseudospin operators and pseudospin Hamiltonian~\cite{ARIMA1969} suitable for the treatment of multipolar interactions~\cite{Pryce1950, Bragam1951, Chibotaru2013},

In the pseudospin picture, spin and orbital degrees of freedom are not independent, and the operator set is formed by a unit multipole (tensor) operator $T^Q_{K}(J)$~\cite{Sakurai, Santini2009,Savrasov2014} (here $J$ is the angular moment, $K$ the rank, and $Q=-K,...,K$). In a general form, a multipolar exchange Hamiltonian can be written as~\cite{Santini2009,Savrasov2014}
\begin{equation}
H =\sum_{ij}\sum_{KQ}C_{K_{i}K_{j}}^{Q{i}Q_{j}}T_{K_{i}}^{Q_{i}}T_{K_{j}}^{Q_{j}},
\end{equation}
where  $i$,$j$ are the site indexes, and $C_{K_{i}K_{j}}^{Q{i}Q_{j}}$  are the coupling constants describing how the energy of the system changes as a result of variation of the two multipole moments $T_{K_{i}}^{Q_{i}}$ and $T_{K_{j}}^{Q_{j}}$.

UO$_2$ adopts a noncollinear 3$\kk$-ordered magnetic configuration, and the two-electrons ($f^2$) ground state of a uranium ion is a $\Gamma_5$ triplet, corresponding to the effective spin (pseudospin) $\tilde{S}=1$~\cite{Mironov2003}. The $\Gamma_5$ ground state is associated with cooperative quadrupolar interactions that cannot be accounted for by using an $S=1/2$ Heisenberg model~\cite{Savrasov2014,GIANNOZZI198775,Mironov2003,Carretta2010,Caciuffo2011}, but can be modeled by means of a suitable pseudospin Hamiltonian.

\begin{figure}
\includegraphics[width=0.3\textwidth,clip]{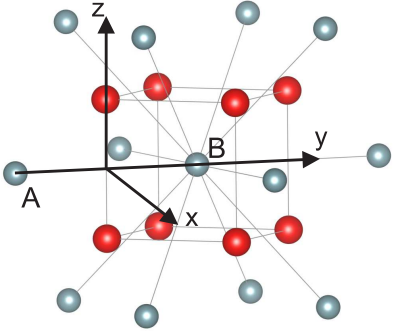}
\caption{Spin quantization axes $x$, $y$, and $z$ and the geometry of the exchange-coupled pair for neighboring  U$^{4+}$ exchange ions. There are six inequivalent exchange-coupled pairs AB per a magnetic unit cell. }
\label{fig:coordinates}
\end{figure}

Below, we adopt the multipolar spin Hamiltonian derived by Mironov \emph{et al.}~\cite{Mironov2003}, describing superexchange interactions between neighboring U$^{4+}$ ions in the 5$f^2$ configuration. The general form of the Mironov exchange Hamiltonian is
\begin{equation}\label{eq:Htot}
\begin{split}
H=A_0+H_1+H_2+H_3+H_4,
\end{split}
\end{equation}
where $A_0$ is a spin-independent parameter, whereas the remaining terms account for various types of spin interactions, which can be written using the conventional spin variables as
\begin{align}
H_1 &=\begin{aligned}[t]
    & D[(S^z_{\rm A})^2+(S^z_{\rm B})^2] \\
    &+E[(S^x_{\rm A})^2-(S^y_{\rm A})^2+(S^x_{\rm B})^2-(S^y_{\rm B})^2],
\end{aligned}
\label{eqnh1} \\[\jot]
H_2 &=\begin{aligned}[t]
    &J_x S^x_{\rm A} S^x_{\rm B} + J_y S^y_{\rm A} S^y_{\rm B} + J_z S^z_{\rm A} S^z_{\rm B},
    \end{aligned}
\label{eqnh2} \\[\jot]
H_3 &= \begin{aligned}[t]
    & j_1 S^x_{\rm A} S^x_{\rm B} [ (S^z_{\rm A})^2+(S^z_{\rm B})^2]  +2 j_1  S^y_{\rm A} S^y_{\rm B} S^z_{\rm A} S^z_{\rm B} \\
    &+ j_2 S^y_{\rm A} S^y_{\rm B} [ (S^z_{\rm A})^2+(S^z_{\rm B})^2]  +2 j_2 S^x_{\rm A} S^x_{\rm B} S^z_{\rm A} S^z_{\rm B},
     \end{aligned}
     \label{eqnh3} \\[\jot]
H_4 &=\begin{aligned}[t]
    & q_1 O^{\rm (1)}_{\rm A} O^{\rm (1)}_{\rm B} +q_2 O^{\rm (2)}_{\rm A}  O^{\rm (2)}_{\rm B}  + q_3 O^{\rm (3)}_{\rm A} O^{\rm (3)}_{\rm B} \\
&+q_4 [O^{\rm (1)}_{\rm A} O^{\rm (2)}_{\rm B}+O^{\rm (2)}_{\rm A} O^{\rm (1)}_{\rm B}].
      \end{aligned}
     \label{eqnh4}
\end{align}
Here, $H_1$ is a single-spin term, quadratic in the spin components and accounting for the zero field splitting (ZFS) dipolar interactions; the ZFS parameters $D$ and $E$ describe the axial and transversal components of magnetic dipole-dipole (DD) interaction, respectively.
$H_2$ is bilinear in spins and describes spin exchange interactions, parameterized by $J_x$, $J_y$ and $J_z$.
Term $H_3$ describes four-spin exchange interactions with $j$ as the corresponding coupling constant.
Finally, $H_4$ accounts for biquadratic quadrupole-quadrupole (QQ) interactions, where $O^{\rm (n)}_{\rm A,B}$ are the components of the quadrupole operator, specifically:
\begin{eqnarray}
O^{\rm (1)}_k &=& (S^z_k)^2-S(S+1)/3, \\
O^{\rm (2)}_k &=&  (S^x_k)^2-(S^y_k)^2, \\
O^{\rm (3)}_k &=& S^x_k S^y_k +S^y_k S^x_k,
\end{eqnarray}
where $k=$A, B.
Labels A and B refer to the nearest neighbor uranium ions (see Fig.~\ref{fig:coordinates} for details). There are four inequivalent sites in a magnetic unit cell of UO$_2$ [U1--U4; see Fig.~\ref{fig:theta}(a)] producing six inequivalent nearest neighbor AB pairs:
A=U1, B=U2; A=U1, B=U3; A=U1, B=U4; A=U2, B=U3; A=U2, B=U4; A=U3, B=U4.
$S_{\rm A}$ and $S_{\rm B}$ are the two spins forming a distinct inequivalent AB pair, and
$x$, $y$, and $z$ are the local quantization axes illustrated in Fig.~\ref{fig:coordinates}.
Cartesian components of spins $S_{\rm A}$ and $S_{\rm B}$ in the local quantization axis representation are given in the Supplemental Material~\cite{SM}.

Performing fully \emph{ab initio} evaluation of the superexchange parameters is difficult \cite{Savrasov2014,Savrasov2014a}. Depending on the definition of tensor operators, slightly different forms of superexchange coupling have been proposed \cite{Mironov2003,Savrasov2014,GIANNOZZI198775,Carretta2010}, impeding accurate quantitative comparison.

In formulating the effective pseudospin Hamiltonian we follow  Mironov~\cite{Mironov2003}. To estimate effective magnetic interactions, Mironov {\it et al.}  used  a second-order perturbation approach, treating free-ion and cubic crystal-field parameters and limiting the interaction to the two nearest-neighbor U ions. We estimate the dominant exchange couplings by means of a controlled fitting procedure, involving the mapping of $\Delta E(\theta)$ onto DFT+$U$+SOC total energies using Eq.~(\ref{eq:Htot}).
To achieve this, we have rewritten the four terms entering Eq.~(\ref{eq:Htot}) as functions of the canting angle $\theta$, replacing components of spins by their explicit expressions in terms of local Cartesian components and arriving at the total magnetic energy expressed as a function of $\theta$.
After some algebra, we find that $\Delta E(\theta)$ has the form,
\begin{equation}\label{eq:Htall}
\begin{split}
& \Delta E(\theta)=B_0+ B_1 \cos(\theta) +B_2 \cos(2 \theta)+B_3\cos(3\theta)  \\
 &\  +B_4\cos(4\theta)  +C_1 \sin(\theta)+C_2\sin(2 \theta) + C_4 \sin(4 \theta),
\end{split}
\end{equation}
where the coefficients are given in terms of the 11 superexchange parameters entering the Mironov Hamiltonian ($D$, $E$, $J_x$, $J_y$, $J_z$, $j_1$, $j_2$, $q_1$, $q_2$, $q_3$, and $q_4$):
\begin{align}
B_0 &=\begin{aligned}[t]
&6 D + 3/2 J_x + 3/2 J_y + 3 J_z +3 j_1 + 3 j_2 \\
&+  3 q_1 + 3/2 q_2 + 3 q_4 - 2 q_1 S + ( 4 D - 2 E \\
& - 2 j_1 - 2 j_2 - 3/2 J_x - 5/2 J_y + 2 J_z  \\
& +14/3 q_1 -  4 q_2 -4 q_3 - q_4) S^2 + 2/3 q_4 S^3 \\
&+ 1/24 (11 q_1+ 42 q_2 + 33 q_3 + 4 q_4 \\
& -42 j_1- 42 j_2) S^4,
\end{aligned}
\label{eq:A0} \\[\jot]
B_1&=\begin{aligned}[t]
&(4 D + 2 J_z + 2 j_1+ 2 j_2 +4 q_1 + 2 q_4 )S\\
&- 4/3 q_1 S^2 -(j_1+  j_2 - 5/3 q_1 +3 q_4 ) S^3,
    \end{aligned}
\label{eq:B1} \\[\jot]
B_2 &= \begin{aligned}[t]
& (2 E - 1/2 J_x + 1/2 J_y + q_4 )S^2- 2/3  q_4 S^3 \\
&+ ( j_1  + j_2  + 1/2 q_1  - q_2 + 3/2 q_3  - 2/3 q_4 )S^4,
     \end{aligned}
     \label{eq:B2} \\[\jot]
B_3 &= \begin{aligned}[t]
& (j_1 + j_2 + q_1 + 3 q_4) S^3,
     \end{aligned}
     \label{eq:B3} \\[\jot]
B_4 &= \begin{aligned}[t]
&  1/8 (6 j_1 + 6 j_2 + 3 q_1 - 6 q_2 + 9 q_3 + 4 q_4) S^4,
     \end{aligned}
     \label{eq:B4} \\[\jot]
C_1 &= \begin{aligned}[t]
&  (4 D + 2j_1 + 2j_2 + 2J_z + 4 q_1 + 2q_4)\sqrt{2}S \\
& - 4\sqrt{2}/3 q_1 S^2 + \sqrt{2}/6 (q_1-15 j_1 -15j_2 ) S^3,
     \end{aligned}
     \label{eq:C1} \\[\jot]
C_2 &= \begin{aligned}[t]
& (-4 E + J_x - J_y -2 q_4)\sqrt{2}S^2 \\
&+ 4\sqrt{2}/3 q_4 S^3+\sqrt{2}/3 q_4 S^4,
     \end{aligned}
     \label{eq:C2} \\[\jot]
C_4 &=\begin{aligned}[t]
    &\sqrt{2}/2 q_4 S^4.
      \end{aligned}
     \label{eq:C4}
\end{align}

\subsection{Evaluating the strength of spin-orbit coupling}
We conclude this section with an estimate of the strength of SOC in UO$_2$. To produce this estimate, we relate the relativistic total energies obtained from first principles calculations to the relativistic atomic Hamiltonian for $f$ orbitals:
\begin{eqnarray}\label{SOC_H}
H_{\rm SOC} = \lambda\,\,  \textbf{L} \cdot \textbf{S},
\end{eqnarray}
where $\lambda$ defines the strength of SOC. Using the 14  $f$ ($l=3$) spinors as a basis,
we write the atomic Hamiltonian $H_{\rm SOC}$ as a (14$\times$14) matrix, see
Eq.~(\ref{mat:occ}) in the appendix.

The diagonalization of this matrix (\ref{mat:occ}) yields the following eigenvalues:
$-2\lambda$, $-2\lambda$, $-2\lambda$, $-2\lambda$, $-2\lambda$, $-2\lambda$, $3\lambda/2$, $3\lambda/2$, $3\lambda/2$, $3\lambda/2$, $3\lambda/2$, $3\lambda/2$, $3\lambda/2$, $3\lambda/2$. From these eigenvalues, we extract the SOC contribution to the total energy $\Delta E_{\rm soc}$ by  considering either the SOC-induced splitting ($\frac{7}{2}\lambda$) or the energy contribution arising from the occupied states. We have followed the latter route, as in this case a suitable mapping can be constructed between the atomic limit ($H_{\rm SOC}$) and an {\it ab initio} calculation. Noting that U$^{4+}$ ions in UO$_2$ are in the $5f^2$ electronic configuration,  the two electrons occupy the lowest two eigenvalues ($-2\lambda$) resulting in $\Delta E_{\rm soc}=-4 \lambda$.
An estimate of $\Delta E_{\rm soc}$ can be obtained from the DFT total energy difference between a relativistic (with SOC) and a nonrelativistic calculation (no SOC), i.e., $\Delta E_{\rm soc}=E_{\rm SOC}-E_{\rm no SOC}$.
To exclude the spurious energy contributions arising from differences in the electronic ground states (insulating vs. metallic), this estimate was obtained using $U=0$ and the 3\textbf{k}
spin-ordered configuration.
In this limit, and both reference states (with and without SOC) are metallic. On the other hand, a DFT+$U$+SOC calculation delivers an insulating solution, where DFT+$U$ (without SOC) stabilizes an insulating ground state. Exploring this would involve terms other than SOC to the energy balance affecting the evaluation of the SOC energy.  Using the above approach ($U$=0), we find that $\Delta E_{\rm soc}=-2.90$ eV per uranium ion, corresponding to $\lambda=-\Delta E_{\rm soc}/4$=0.73 eV.
We should mention that in the metallic solution, the values of spin and orbital moments are greatly reduced with respect to those found in the DFT+$U$+SOC ground state (referred to in Table~\ref{table:moments}), specifically $m_s$ $\approx$ 1 $\mu_B$ and $m_o$ $\approx$ $1.5~\mu_B$.

\begin{table}
\caption{Values of spin ($m_s$), orbital ($m_o$), and total ($m_t$) moments (in $\mu_B$) and the electronic band gap $E_g$ (in eV) corresponding to several different values of parameter $U$ (in eV). The experimentally observed band gap and the local total magnetic moment on uranium ions are  $\approx$ 2.0~eV~\cite{Schoenes1980} and 1.74~$\mu_B$~\cite{Frazer1965}, respectively.}
\begin{ruledtabular}
\begin{tabular}{lcccc}
     & $U$=3.16 &  $U$=3.46 & $U$=3.96 & $U$=4.50 \\
\hline
$m_s$    &  -1.52   &    -1.53  &   -1.54   &    -1.54 \\
$m_o$    &   3.18   &     3.19  &    3.21   &     3.22 \\
$m_t$    &   1.66   &     1.66  &    1.67   &     1.68 \\
$E_g$    &   1.91   &     2.11  &    2.44   &     2.78
 \end{tabular}
\end{ruledtabular}
\label{table:moments}
\end{table}

\begin{figure}
\includegraphics[width=0.4\textwidth,clip]{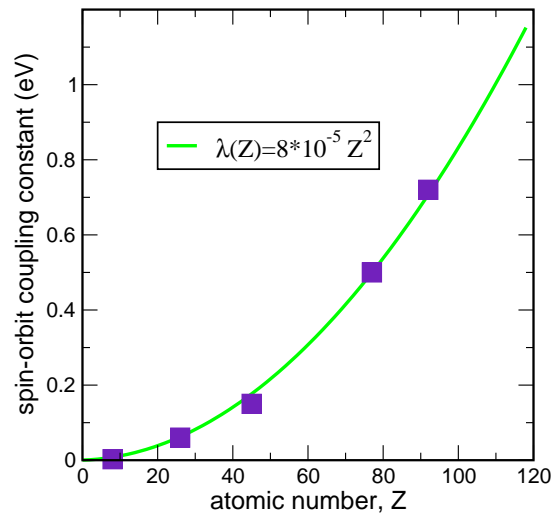}
\caption{Variation of the spin-orbit coupling constant $\lambda$ as a function of nuclear charge $Z$ compared with literature data for oxygen \cite{Tinkham1955}, iron \cite{Desjonqueres2007}, rhodium \cite{Chikara2015}, iridium \cite{IrSOC,PhysRevLett.101.076402}, and uranium (this work).}
\label{fig:SOC_constant_fitted}
\end{figure}

Further support for this large value of $\lambda$ comes from an approximate scaling of the magnitude of SOC at atomic level, where it is known that the SOC parameter $\lambda$ scales as $\sim Z^2$, where  $Z$ is the atomic number~\cite{LandauLifshitzQM}. By rescaling the SOC strength of iridium (0.5 eV)~\cite{IrSOC} with the relative nuclear charge of Ir ($Z_{\rm Ir}=77$) and U ($Z_{\rm U}=92$) we find
\[
\lambda_{\rm U} \approx \lambda_{\rm Ir} (Z_{\rm U}/Z_{\rm Ir})^2 = 0.5~{\rm eV} \times (92/77)^2 = 0.71~{\rm eV},
\]
in good agreement with the {\it ab initio} estimate. An interpolation formula $\lambda (Z)=8\cdot 10^{-5}Z^2$ (eV) appears to provide a good match to the data available from literature on the strength of SOC in relatively heavy elements (see Fig.~\ref{fig:SOC_constant_fitted}), where spin-orbit interaction plays a significant part.

\section{Results and Discussion}

\begin{figure}
\includegraphics[width=0.45\textwidth,clip]{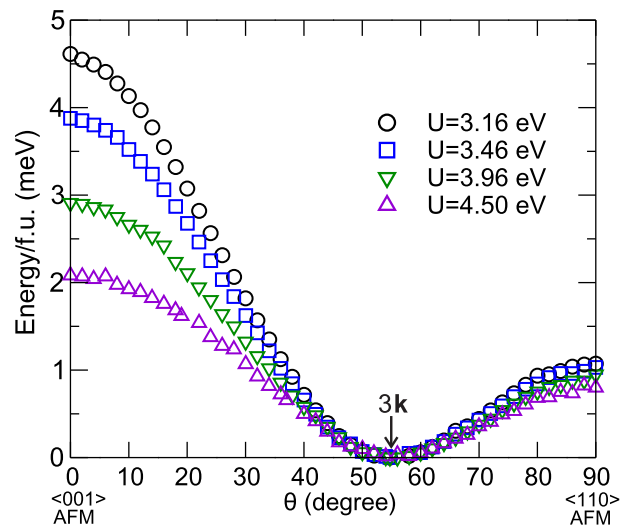}
\caption{Total energy as a function of the canting angle $\theta$ computed for several different values of parameter $U$.}
\label{fig:E_theta}
\end{figure}

The large SOC in UO$_2$ is responsible for the formation of a $\Gamma_5$ triplet described by an effective pseudospin $\tilde{S}=1$~\cite{Mironov2003} state, where the spin and orbital moments are ordered in a 3$\kk$ magnetic structure~\cite{BURLET1986121,Amoretti1989}; see Figs.~\ref{fig:theta}(a) and~\ref{fig:theta}(b). Moreover, the various types of (multipolar) superexchange interactions acting in the 3$\kk$ magnetic configuration are coupled with the cooperative Jahn-Teller effect, manifested by a distortion of the oxygen cage around the U$^{4+}$ ions~\cite{Allen1968b,Santini2009,GIANNOZZI198775,Jaime2017}.
The computational verification of these experimental observations and their interpretation on a quantum level is a difficult task due to a variety of factors: (i) magnetic noncollinearity, (ii) self-interaction acting in the U-$f$ manifold, and (iii) existence of multiple local minima in a narrow energy interval~\cite{Laskowski2004, Dorado2009, Zhou2011}. As was noted above, a combination of fully relativistic and magnetically constrained DFT+$U$ with the adiabatic evolution of the occupation matrix is able to predict the ground state of UO$_2$ [Fig.~\ref{fig:theta}(c)] and should help decipher the subtleties of electronic and magnetic effects in UO$_2$.

To gain insight into the nature of magnetic interactions, we compute the magnetic energy curves similar to the one shown in Fig.~\ref{fig:theta}, but this time we perform the calculations for several different values of parameter $U$ used in the DFT+$U$ formalism. The curves shown in Fig.~\ref{fig:E_theta} suggest that the noncollinear 3$\kk$ ordering remains the lowest energy state for any value of $U$, but the energy difference between the 3$\kk$ phase and the competing AFM collinear phases $\lC 001 \rC$ and $\lC 110 \rC$, illustrated in Fig.~\ref{fig:models}, depends sensitively on the choice of $U$. As the value of parameter $U$ increases, the relative stability of the 3$\kk$ state decreases, and it becomes progressively less energetically costly to rotate the spins. This implies that the value of the magnetic exchange interactions is also sensitive to the choice of $U$. We shall discuss this later in the section.

\begin{table}
\caption{Relative energies (in meV/f.u.) of the $\langle 001\rangle $-AFM and $\langle 110\rangle $-AFM with respect to 3$\kk$ phases as a function of the Jahn-Teller distortion ($\delta_{JT}$, \AA) and the strength of SOC, rescaled to a half of the self-consistent value $\lambda=0.73$ eV. The phases with $\delta_{JT}=0$ and $\delta_{JT}=0.003$ are almost degenerate in energy within 10$^{-5}$ eV/f.u., whereas the experimental structure ($\delta_{JT}=0.014$) is 0.84 meV/f.u. less stable than the self-consistently optimized one (see Supplemental Material~\cite{SM}). All the data given in the table were computed for $U$=3.46~eV.}
\begin{ruledtabular}
\begin{tabular}{lcc}
JT effect & $E_{\langle 001\rangle \rm {-AFM}}-E_{3\kk}$ & $E_{\langle 110\rangle \rm {-AFM}}-E_{3\kk}$ \\
$\delta_{JT}=0$  &  3.86     & 1.01  \\
$\delta_{JT}=0.003$  & 3.87    & 1.03   \\
$\delta_{JT}=0.014$  & 3.88   &  0.94   \\
\hline\hline
SOC strength & $E_{\langle 001\rangle \rm {-AFM}}-E_{3\kk}$ &  $E_{\langle 110\rangle \rm {-AFM}}-E_{3\kk}$ \\
$\lambda$  &  3.87  &  1.03 \\
0.5$\lambda$  & -8.78  & -11.28
%
\end{tabular}
\end{ruledtabular}
\label{table:socmag}
\end{table}

The role of parameter $U$ is also reflected in the fundamental  electronic and magnetic properties of the 3$\kk$ ground state. Table ~\ref{table:moments} gives values of the spin moment $m_s$, the orbital moment $m_o$, and the insulating gap $E_g$ computed for $U$=3.16, 3.46, 3.96 and 4.50~eV. While the moments are only marginally affected by the choice of $U$ (and all of them compare well with the observed total moment of  1.74~$\mu_B$~\cite{Frazer1965}), the band gap varies significantly from 1.91~eV to 2.78~eV. The best agreement with experiment ($E_g$$\approx$2.0 eV~\cite{Schoenes1980} ) is found for the relatively small $U$, in agreement with the first principles estimate of the Coulomb interaction parameters based on cRPA (see Sec.~\ref{sec:crpa}), and also in agreement with results derived from a recent fitting analysis~\cite{Pegg2017}.
We also note that even though the value of parameter $J$ in UO$_2$ is not very large, 0.30~eV, reducing $U$ by 0.3 eV reduces the band gap by about 10\% ($E_g$=1.91~eV for $U=U^{\rm cRPA}-J^{\rm cRPA}$=3.16~eV and  $E_g$=2.11~eV for $U=U^{\rm cRPA}$=3.46~eV, both fairly close to the observed band gap of approximately 2.0~eV).

\begin{table}
\caption{Magnetic moments of uranium ions computed for the $\langle 001\rangle $-AFM, $\langle 110\rangle $-AFM and 3$\kk$ magnetic configurations. All the values were computed assuming $U$=3.46~eV.}
\begin{ruledtabular}
\begin{tabular}{lccc}
         &   $\langle 001\rangle $-AFM &  $\langle 110\rangle $-AFM  & 3$\kk$ \\
\hline
$m_s$    & -1.55 & -1.51 & -1.53 \\
$m_o$    & 3.26  &  3.24 &  3.19\\
$m_t$    & 1.70  &  1.73 &  1.66
\end{tabular}
\end{ruledtabular}
\label{table:moments2}
\end{table}

Next, we examine the quantum mechanism responsible for the onset of magnetic 3$\kk$ ordering. We remind the reader that in a Jahn-Teller-active material with strong SOC, the Jahn-Teller instability and exchange interactions are antagonists since the JT effect tends to stabilize states with quenched orbital momentum whereas SOC tends to maximize the orbital momentum~\cite{Goodenough}. This conclusion is generally valid if the crystal field is large, but in UO$_2$ the strength of SOC is very large ($\approx$ 0.73~eV according to the estimate above) exceeding the energy scale of crystal-field excitations (150-180 meV~\cite{Amoretti1989}), and therefore the spin-orbit interaction can be safely considered as the dominant energy scale and the leading factor stabilizing the 3$\kk$ state. To verify this hypothesis, we have calculated the total energy of 1$\kk$, 2$\kk$, and 3$\kk$ magnetically ordered states as a function of strength of the JT effect, switching it on and off.
In the on mode we have tested both the self-consistently derived (0.003~\AA) and the experimentally observed (0.014~\AA~\cite{PhysRevLett.35.1770}) values of JT-displacements, and examined two values of SOC, the full SOC strength $\lambda=0.73$~eV and half the SOC strength $\lambda=0.36$~eV. The data, given in Table~\ref{table:socmag}, show that the JT effect has virtually no effect on the relative energies of magnetic configurations. The energy landscape and the energy difference between the 1$\kk$, 2\textbf{k} and 3\textbf{k} states are not affected by the strength of JT distortion and remain essentially unchanged (see Table~\ref{table:socmag} and Supplemental Material~\cite{SM}). This unequivocally demonstrates that the JT effect is not the mechanism that drives the system towards the 3$\kk$ ground state~\cite{sk2018}.
On the other hand, rescaling the SOC strength to half the original value ($\lambda=0.36$~eV) causes a huge energy change, favorable for both the $\langle 001\rangle $-AFM and $\langle 110\rangle $-AFM configurations, where the latter as a result of halving the SOC value becomes the most favorable one by more than 11 meV/f.u.
This provides a clear indication that SOC is the major driving force responsible for the stabilization of the 3$\kk$ state; reducing the SOC strength leads to the over-stabilization of collinear magnetic structures.

\begin{figure}
\includegraphics[width=0.47\textwidth,clip]{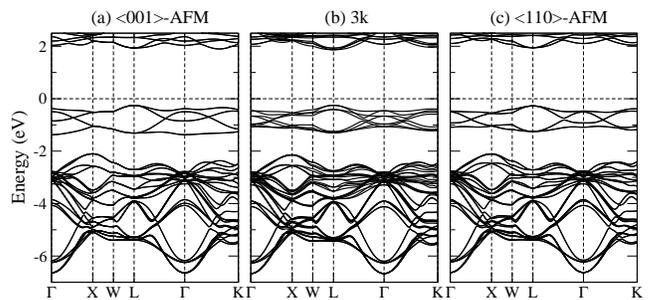}
\caption{Band structures of UO$_2$ computed for the $\langle 001\rangle $-AFM,  3$\kk$ and $\langle 110\rangle $-AFM ordered configurations assuming $U$=3.46~eV.}
\label{fig:bands}
\end{figure}

\begin{table}
\caption{Eigenvalues of the 14$\times$14  occupation matrix of the $f$ manifold of uranium ions in UO$_2$ computed
for the $\lC 001 \rC$-AFM, 3$\kk$ and $\lC 110 \rC$-AFM spin configurations. All the values were computed for $U$=3.46~eV.}
\begin{ruledtabular}
\begin{tabular}{ccc}
  $\lC 001 \rC$-AFM &  3$\kk$ & $\lC 110 \rC$-AFM \\
\hline
0.0268 & 0.0271 & 0.0273 \\
0.0274 & 0.0282 & 0.0275 \\
0.0296 & 0.0288 & 0.0296 \\
0.0316 & 0.0341 & 0.0333 \\
0.0348 & 0.0356 & 0.0359 \\
0.0390 & 0.0365 & 0.0362 \\
0.0397 & 0.0366 & 0.0375 \\
0.0398 & 0.0384 & 0.0391 \\
0.0454 & 0.0488 & 0.0477 \\
0.0501 & 0.0508 & 0.0513 \\
0.1233 & 0.1238 & 0.1238 \\
0.1393 & 0.1407 & 0.1404 \\
0.9852 & 0.9846 & 0.9846 \\
0.9888 & 0.9858 & 0.9860
\end{tabular}
\end{ruledtabular}
\label{table:occupations}
\end{table}

To discern how SOC stabilizes the 3$\kk$ ordering of moments, we have explored the differences between electronic and magnetic properties of these three phases. Surprisingly, there are only marginal changes in the magnitude of spin and orbital moments (see Table~\ref{table:moments2}), in the band structure (see Fig.~\ref{fig:bands}), as well as in the occupation numbers of states in the $f$ manifold (see Table~\ref{table:occupations}). A closer inspection of the band structure shows that even though the overall bonding picture in all the three magnetically ordered configurations is almost identical (including the size of the band gap), the $f$-manifold in the 3$\kk$ phase exhibits larger SOC-induced splitting, which causes a change in the band topology.

\begin{figure*}
\includegraphics[width=0.90\textwidth,clip]{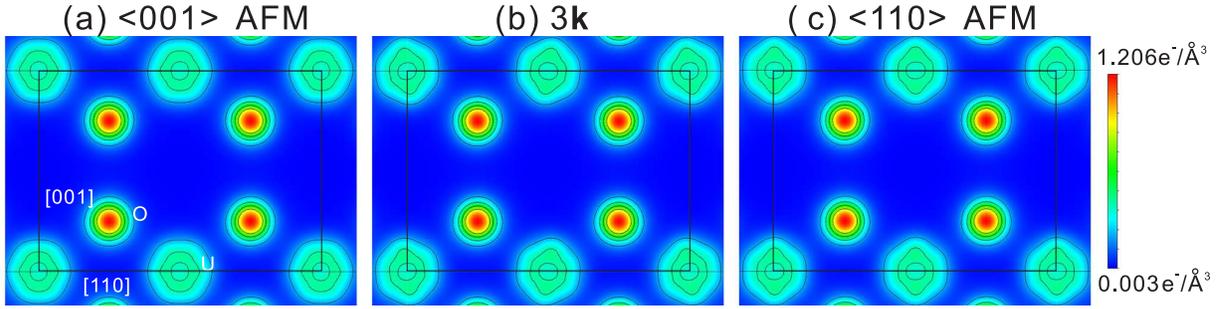}
\caption{Charge densities of UO$_2$ in a (110) plane computed for the $\lC 001 \rC$-AFM,  3$\kk$ and $\lC 110 \rC$-AFM spin ordered configurations. All the three plots were computed assuming that $U$=3.46~eV.}
\label{fig:charge}
\end{figure*}

To better understand the significance of this change in the topology of band structure, in Fig.~\ref{fig:charge} we plot charge-density isosurfaces for the occupied $f$ orbitals in the energy interval (-2, 0) eV, projected onto a (110) plane containing both uranium and oxygen ions. The results show that the 3$\kk$ configuration is the only spin arrangement exhibiting a visible orbital anisotropy at the uranium sites, associated with the canted ordering of $f$ orbitals. The $f$-orbitals are rotated towards the nearest oxygen sites, following the same chessboard configuration of the 3$\kk$ spin ordering as that shown in Fig.~\ref{fig:theta}(a) and (b). Remarkably, the effect of SOC, critical to the stabilization of the 3$\kk$ phase, is manifested primarily in the shape of $f$-orbitals rather than in the total orbital occupation, as illustrated in Table~\ref{table:occupations}. The energy required to stabilize the 3$\kk$ state over the $\langle 001\rangle $-AFM and $\langle 110\rangle $-AFM configurations is gained from a SOC-induced rotation of occupancies of particular orbitals, which follows the rotation of the local spin moments and enables constructive interaction with the oxygen electronic states.

\begin{table*}
\caption{Magnetic coupling parameters (meV) estimated by Mironov~\cite{Mironov2003} shown together with the fitted values of the dominant QQ terms obtained by mapping the DFT+$U^{\rm cRPA}$+SOC energies onto the extended Mironov's Hamiltonian. $\chi_R^2$ serves as an indication of the quality of the fit in terms of the reduced chi-squared test.}
\begin{ruledtabular}
\begin{tabular}{lrrrrrrrrrrrrr}
         &  $D$ & $E$ & $J_x$ & $J_y$ & $J_z$ & $j_1$ & $j_2$ & $q_1$ & $q_2$ & $q_3$ & $q_4$  & $\chi_R^2$ \\
\hline
Mironov~\cite{Mironov2003} & $-$0.57 & 0.64 & 1.82 & 2.74 &2.33 &$-$0.04 & 0.05 & 0.22 &0.32 & 0.07 &$-$0.02  &  \\
Fit (QQ) &  &  & & & & &  & 7.18 & 1.94 & & & 0.9989
\end{tabular}
\end{ruledtabular}
\label{table:fit_para}
\end{table*}

\begin{figure}
\includegraphics[width=0.43\textwidth,clip]{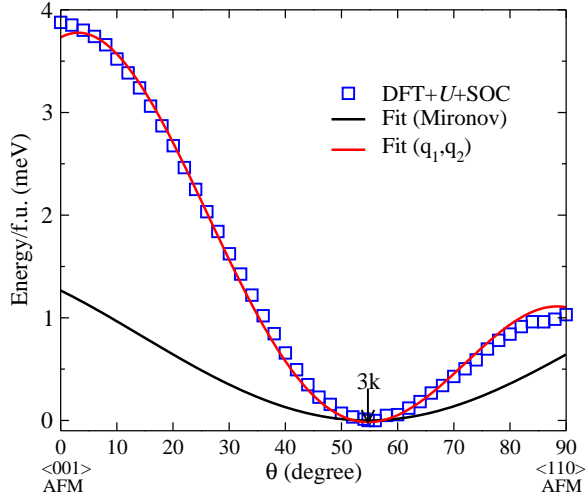}
\caption{Comparison between the calculated (DFT+$U^{\rm cRPA}$+SOC) and fitted (out fit using Mironov's parameters) magnetic canting energies $\Delta E(\theta)$. In our fit we employed  an extension of the Mironov model to all the U-U interactions and optimized the fit with respect to the dominant quadrupolar terms $q_1$ and $q_2$, by keeping all the other terms fixed to the corresponding values obtained by Mironov~\cite{Mironov2003} (see Table~\ref{table:fit_para}).}
\label{fig:fit}
\end{figure}

Having established the DFT+$U^{\rm cRPA}$+SOC as a suitable theory for the ground state electronic properties of UO$_2$, highlighting the significance of SOC in this compound, we are now ready to proceed to the analysis of superexchange spin interaction mechanisms, to deduce information about the quantum origin of the 3$\kk$ state. As we noted in the computational section, this can be done by fitting the magnetic energy computed using {\it ab initio} methods, to the multipolar Hamiltonian (\ref{eq:Htall}). The presence of eleven parameters in the multipolar Hamiltonian clearly poses a well known problem for the multi-parameter fitting procedure~\cite{neumann}. To handle this complication, we rely on the analysis by Mironov \emph{et al.}~\cite{Mironov2003} that can be summarized as follows. First, we note that using the values of parameters evaluated by Mironov (collected in Table~\ref{table:fit_para}) in the pseudospin Hamiltonian, already leads to an overall fairly good account of the first principles magnetic energy, as illustrated graphically in Fig.~\ref{fig:fit}. Even though the two curves do not match well, Mironov's parameters predict the correct position of the energy minimum, located at the 3$\kk$ position, suggesting that all the relevant magnetic coupling terms are correctly included in the theoretical treatment. However, Mironov's parameters deliver a curve that varies over a significantly narrower energy interval than the curve derived from the first principles data, and as a result the relative stability of the 3$\kk$ state with respect to the collinear 1\textbf{k} and 2\textbf{k} states is underestimated by about 50\%.

The exchange parameters in Mironov's model fall into four different categories, namely (i)  single-spin parameters $D$ and $E$, (ii)  bilinear  parameters $J_x$, $J_y$ and $J_z$, (iii) parameters describing the four-spin terms $j_1$ and $j_2$ and (iv) parameters of biquadratic interactions $q_1$, $q_2$, $q_3$, and $q_4$.
The accuracy of Mironov's approach can be improved by noting that, according to the calculations by Savrasov and coworkers, the strength of quadrupolar (QQ) interactions computed by Mironov is underestimated by an order of magnitude~\cite{Savrasov2014}. Following this  argument, we have fitted $\Delta E(\theta)$ by varying the two largest quadrupolar terms ($q_1$, $q_2$) only, and keeping all the other superexchange parameters fixed to the original Mironov's values. The resulting curve is in excellent agreement with first principles energies ($\chi_R^2$=0.9989, see Fig.~\ref{fig:fit}), and this improvement is associated with a very large increase of the magnitude of the quadrupolar terms, approximately by an order of magnitude ($q_1$+$q_2$)/2=4.56~meV, see Table~\ref{table:fit_para}. However, this is in very good agreement with the earlier DFT+$U$ data (3.1~meV~\cite{Savrasov2014}) and the values extracted from experimental spin-wave spectra (1.9~meV)~\cite{Caciuffo2011}.
As expected, the values of $q_1$ and $q_2$ are sensitive to the choice of parameter $U$, see Table~\ref{table:q1q2}. The strength of the QQ interaction increases as a function of $U$, this in particular applies to the anisotropic biquadratic interaction $q_2$. Nevertheless, the resulting values do not depend on the Jahn-Teller distortions; using the undistorted cubic phase one obtains essentially the same values of parameters, further demonstrating the fairly negligible role played by the JT effect in stabilizing the 3$\kk$ noncollinear state.

\begin{table}
\caption{Fitted quadrupolar parameters $q_1$ and $q_2$ (and their average, in meV) as a function of $U$.}
\begin{ruledtabular}
\begin{tabular}{lccc}
          &   $q_1$    &  $q_2$  &  ($q_1$+$q_2$)/2\\
 \hline
$U$ = 3.16 eV  &   6.55  &  0.36 & 3.46\\
$U$ = 3.46 eV &   7.18  &  1.94 & 4.56 \\
$U$ = 3.96 eV &   7.64  &  3.77 & 5.71 \\
$U$ = 4.50 eV  &   7.81  &  5.19 & 6.50
\end{tabular}
\end{ruledtabular}
\label{table:q1q2}
\end{table}

\begin{figure}
\includegraphics[width=0.52\textwidth,clip]{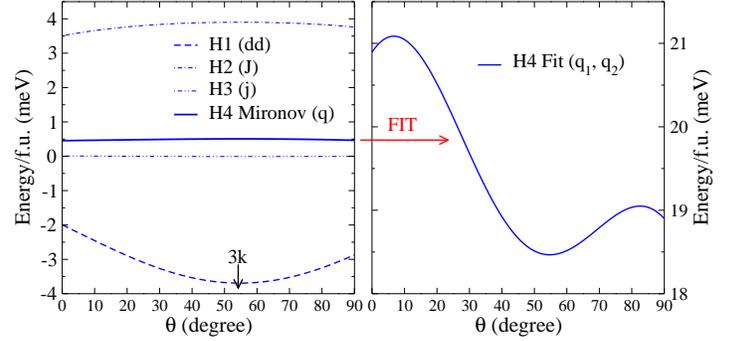}
\caption{Decomposition of the magnetic energy $\Delta E(\theta)$ into four components $H_1$ [DD interaction, $D$ \& $E$, Eq.~(\ref{eqnh1})], $H_2$ [bilinear exchange, $J$'s, Eq.~(\ref{eqnh1})], $H_3$ [four-spin exchange, $j$'s, Eq.~(\ref{eqnh3})], and $H_4$ [quadrupolar, $q$'s, Eq.(~\ref{eqnh4}]. The right panel shows the difference between the quadrupolar term derived directly from Mironov's data and our fitted curve.}
\label{fig:decomposition}
\end{figure}

We conclude the discussion of magnetic properties of UO$_2$ by analyzing the individual contributions of various types of superexchange mechanisms to the stabilization of the 3$\kk$ state. As was noted in the section on computational methods, the total magnetic Hamiltonian is expressed as a sum of four terms, each corresponding to a specific type of superexchange interaction: $H_1$ accounts for the DD interaction [Eq.~(\ref{eqnh1})], $H_2$ describes the bilinear exchange [Eq.~(\ref{eqnh2})], $H_3$ represents the four-spin exchange [Eq.~(\ref{eqnh3})], and finally $H_4$ takes care of the quadrupolar coupling [Eq.~(\ref{eqnh4})]. The data, summarized in
Fig.~\ref{fig:decomposition}, clearly show that the formation of the 3$\kk$ state occurs as a result of a concerted action of the DD and octupolar interactions. The contributions of bilinear exchanges ($H_2$, $J$'s) and four spin exchanges ($H_3$, $j$'s) are essentially independent of the canting angle, resulting in the rather flat curves. On the other hand, the DD interactions have a quadratic-like trend with a marked minimum at 3$\kk$ and the fit-corrected quadrupolar term (the right panel of Fig.~\ref{fig:decomposition}) is not only minimum at 3$\kk$, but also correctly describes the energy pathway from the 3$\kk$ to the 1$\kk$ and 2$\kk$ states, following the trend exhibited by the first principles energies (see Fig.~\ref{fig:fit}).

Based on the above results, we can conclude that the onset of the 3$\kk$ state in UO$_2$ is driven, at the quantum level, by a concerted action of the DD and QQ spin interactions. These interactions are active in the undistorted and JT-distorted crystal lattices, clearly indicating that, despite the existing coupling between the spin and lattice degrees of freedom, the JT instabilities do not contribute to the formation of the noncollinear 3$\kk$ ordered magnetic state, which is present also in the undistorted cubic phase.

\section{Summary and Conclusions}

In this study we have parameterized the LSDA+$U$ model for noncollinear magnetic systems and explained the origin of the canted 3$\kk$ state in UO$_2$ by combining several computational methods including the constrained random phase approximation to compute the Coulomb repulsion parameter $U$ and the Hund's coupling parameter $J$, thus rendering the LSDA+$U$+SOC fully \emph{ab initio}, magnetic constraints to explore the dependence of the total energy on the direction of the spin moment, the adiabatic propagation of the occupation matrix to avoid the multiple minima problem in constructing the magnetic energy landscape, and two different effective Hamiltonians to extract from the \emph{ab initio} data the spin-orbit interaction parameter $\lambda$ and the quadrupole-quadrupole exchange interactions.

The outcome of our study is threefold. First, we have derived an invariant orbital- and spin-dependent formalism for the LSDA+$U$ model suitable for noncollinear magnetism  involving spin and orbital contributions, and have shown that the LSDA+$U$ potential and double counting correction depend only on one parameter $U$, and are independent of the Hund coupling parameter $J$. Second, our data suggest that the spin-orbit interaction parameter in UO$_2$ is as large as 0.73~eV, hence explaining many exotic physical phenomena emerging from the intricate interplay between the spin, charge, and orbital degrees of freedom, explicated by the formation of a multipolar magnetic state with tilted orbital ordering in the $f$-orbital manifold.
Finally, we have uncovered the role of dipole-dipole and quadrupole-quadrupole spin interactions in the formation of the noncollinear 3$\kk$ state and ruled out Jahn-Teller distortions as a factor in stabilizing the 3$\kk$ magnetic ordering. The most relevant energy scales defining the properties of UO$_2$ are summarized in Table~\ref{table:sum}.

\begin{table}
\caption{Summary of parameters controlling the magnitude of the relevant energy scales in UO$_2$: $U$ \& $J$ (from cRPA), the SOC strength parameter $\lambda$, the QQ exchange (from fitting to the spin canting {\it ab initio} data), and DD exchange (from Mironov~\cite{Mironov2003}).}
\begin{ruledtabular}
\begin{tabular}{ccccc}
$U^{\rm cRPA}$ & $J^{\rm cRPA}$ & $\lambda$ & QQ   & DD\\
3.46 eV     &   0.3 eV      &   0.73 eV    & 3.46 meV & $\approx$ 0.6 meV
\end{tabular}
\end{ruledtabular}
\label{table:sum}
\end{table}

In addition to elucidating the complexity of various physical scenarios, these results provide a reference for studies of relativistic noncollinear magnetic materials, in particular 5$d$ transition metal oxides, and enable a quantitatively accurate exploration of technologically relevant aspects of UO$_2$ such as spin and orbital magnetic dynamics, the formation and evolution of structural defects and their diffusion. From this perspective, the study shows that an accurate account of fundamental microscopic interactions derived from a direct application of quantum mechanics can provide a quantitative account of physical processes and critical parameters
(see Table~\ref{table:sum}) that can then be used as input for phenomenological schemes describing macroscopic phenomena such as transport and dissipation.

\section*{Acknowledgments}
Part of this work has been carried out within the framework of the EUROfusion Consortium and has received funding from the Euratom research and training programme 2019$-$2020 under Grant Agreements No. 633053 and No. 755039. Also, it has been partially funded by the RCUK Energy Programme (Grant No. EP/P012450/1). The views and opinions expresses herein do not necessarily reflect those of the European Commission. S.L.D. acknowledge support from the Centre for Non-Linear Studies at LANL and would like to thank M. E. A. Coury, A. P. Horsfield and W. M. C. Foulkes for stimulating discussions. P.L. and C.F. acknowledges support from the Austrian Science Fund (FWF). D.A.A. and C.R.S. acknowledge support from the U.S. Department of Energy, Office of Nuclear Energy, Advanced Modeling and Simulation (NEAMS) program. The computational results presented have been achieved using the Vienna Scientific Cluster (VSC).

S.L.D. and P.L. have agreed to state that they contributed equally to this publication.

\appendix*
\setcounter{equation}{0} \renewcommand{\thefigure}{\arabic{equation}}
\setcounter{table}{0} \renewcommand{\thefigure}{\arabic{table}}
\setcounter{figure}{0} \renewcommand{\thefigure}{\arabic{figure}}

\section{SOC matrix in the $f$ spinors}\label{appendix}

Using the 14 $f$ ($l=3$) spinors as a basis in the following order:
$|xyz,\uparrow\rC$, $|x(5x^2-3r^2),\uparrow\rC$,  $|y(5y^2-3r^2),\uparrow\rC$,  $|z(5z^2-3r^2),\uparrow\rC$,
$|x(y^2-z^2),\uparrow\rC$,  $|y(z^2-x^2),\uparrow\rC$,
$|z(x^2-y^2),\uparrow\rC$ (plus the corresponding $\uparrow \Rightarrow \downarrow$ spinors),  we write the atomic SOC Hamiltonian $H_{\rm SOC}$ (\ref{SOC_H}) as a (14$\times$14) matrix~\cite{PhysRevB.79.045107},
\begin{widetext}
\begin{align}
H_{\rm SOC}= \frac{\lambda}{2}
\left(\!
\begin{array}{rrrrrrrrrrrrrrr}
   0 &      0 &      0 &      0 &     0 &      0 &    2 i &      0 &      0 &       0 &       0 &    2 i &       2 &      0 \\
   0 &      0 &  3 i/2 &      0 &     0 &    i t &      0 &      0 &      0 &       0 &    -3/2 &      0 &       0 &      t \\
   0 & -3 i/2 &      0 &      0 &   i t &      0 &      0 &      0 &      0 &       0 &   3 i/2 &      0 &       0 &    i t \\
   0 &      0 &      0 &      0 &     0 &      0 &      0 &      0 &    3/2 &  -3 i/2 &       0 &      t &     i t &      0 \\
   0 &      0 &   -i t &      0 &     0 &   -i/2 &      0 &   -2 i &      0 &       0 &      -t &      0 &       0 &    1/2 \\
   0 &   -i t &      0 &      0 &   i/2 &      0 &      0 &     -2 &      0 &       0 &    -i t &      0 &       0 &   -i/2 \\
-2 i &      0 &      0 &      0 &     0 &      0 &      0 &      0 &     -t &    -i t &       0 &   -1/2 &     i/2 &      0 \\
   0 &      0 &      0 &      0 &   2 i &     -2 &      0 &      0 &      0 &       0 &       0 &      0 &       0 &   -2 i \\
   0 &      0 &      0 &    3/2 &     0 &      0 &     -t &      0 &      0 &  -3 i/2 &       0 &      0 &    -i t &      0 \\
   0 &      0 &      0 &  3 i/2 &     0 &      0 &    i t &      0 &  3 i/2 &       0 &       0 &   -i t &       0 &      0 \\
   0 &   -3/2 & -3 i/2 &      0 &    -t &    i t &      0 &      0 &      0 &       0 &       0 &      0 &       0 &      0 \\
-2 i &      0 &      0 &      t &     0 &      0 &   -1/2 &      0 &      0 &     i t &       0 &      0 &     i/2 &      0 \\
   2 &      0 &      0 &   -i t &     0 &      0 &   -i/2 &      0 &    i t &       0 &       0 &   -i/2 &       0 &      0 \\
   0 &      t &   -i t &      0 &   1/2 &    i/2 &      0 &    2 i &      0 &       0 &       0 &      0 &       0 &      0
\label{mat:occ}
   \end{array}\!\right), \quad  t=\sqrt{15}/2
\end{align}
\end{widetext}

\bibliographystyle{apsrev4-1}
\bibliography{references}

\end{document}